\documentclass{JHEP3}

\pdfoutput=1

\usepackage{graphicx}
\usepackage{amssymb}
\usepackage{amsmath}
\usepackage{url}

\newcommand{\bi}{\begin{itemize}}
\newcommand{\ei}{\end{itemize}}
\newcommand{\be}{\begin{equation}}
\newcommand{\ee}{\end{equation}}

\newcommand{\tev}{\mathrm{\; TeV}}
\title{Measuring the Polarization of Boosted Hadronic Tops}

\author{David Krohn \\ Department of Physics, Princeton University, Princeton, NJ 08544  \\ E-mail:  \email{dkrohn@princeton.edu} }
\author{Jessie Shelton \\ Department of Physics and Astronomy, Rutgers University, Piscataway, NJ 08854 
\\       {\small and} \\
Department of Physics, Yale University, New Haven, CT 06511  \\ E-mail: \email{j.shelton@yale.edu}}
\author{Lian-Tao Wang \\ Department of Physics, Princeton University, Princeton, NJ 08544   \\ E-mail:  \email{lianwang@princeton.edu}}

\abstract{We propose a new technique for measuring the polarization of hadronically decaying boosted top quarks.  In particular, we apply a subjet-based technique to events where the decay products of the top are clustered within a single jet.  The technique requires neither $b$-tagging nor $W$-reconstruction, and does not rely on assumptions about either the top production mechanism or the sources of missing energy in the event.  We include results for various new physics scenarios made with different Monte Carlo generators to demonstrate the robustness of the technique.}

\preprint{RUNHETC-2009-20}

\begin{document}

%%%%%%%%%%%%%%%%%%%%%%%%%%%%%%%%%%%%%%%%%%%%%%%%%%%%%%%%
\section{Introduction}
%%%%%%%%%%%%%%%%%%%%%%%%%%%%%%%%%%%%%%%%%%%%%%%%%%%%%%%%

The top quark, with its large coupling to the Higgs sector, plays an
important role in models of physics beyond the Standard Model.
Indeed, many such models posit the existence of top partner states
(e.g. the stop squark of SUSY~\cite{Dimopoulos:1981zb} and the $T'$ of little Higgs models~\cite{ArkaniHamed:2001nc}) or
otherwise couple the top to new physics in a special way (as with
KK-gluons~\cite{Agashe:2003zs}).  Measuring the couplings of the top to new states is
therefore essential in distinguishing the correct model of physics
beyond the Standard Model.

One especially interesting aspect of these couplings is their
chirality: whether or not they distinguish left- from right-handed
tops.  Fortunately, the large mass of the top, which makes its study
so interesting for electroweak physics, makes it possible to imagine
measuring the chiral couplings of the top directly.  Unlike
the other quarks, the top decays before hadronization, so information about
its spin is transferred to the distributions of its decay
products~\cite{Kane:1991bg}.  On the other hand, the large mass of 
the top also means that in order for the chiral couplings of tops to new physics to
translate into observable top polarization signals, the tops must be significantly boosted,
as chirality only becomes equivalent to helicity in the massless limit.
Boosted tops are therefore a natural and interesting place to
look for polarization signals.

Conventional methods for measuring the polarization of non-boosted tops begin 
by reconstructing the top rest frame and considering the angular distributions 
of its decay products in that frame, and often focus on
the semi-leptonic decay mode, which can be fully reconstructed if the
only missing energy in the event comes from the neutrino.  Such techniques have been
extended to events where the hadronic top is boosted, but the lepton from the
leptonic top decay is still isolated.   This isolated lepton can then be used to
measure the polarization of its parent top, either by reconstructing the $t\bar t$ 
system ~\cite{Agashe:2006hk} or through the shape of the lepton $p_T$ spectrum ~\cite{Lillie:2007yh}.

When the top quark is highly boosted, however, requiring an isolated lepton 
begins to require a significant acceptance price.  
Moreover, while the large spin analyzing
power of the lepton in standard model top decay makes it particularly useful for
top polarization studies, it is also desirable to develop techniques which can measure polarization in
boosted tops without the need for an isolated lepton.  Being able to study
polarization in boosted hadronic tops increases acceptance,
and has the additional feature of {\it flexibility}: unlike leptonic tops, hadronic tops
are fully reconstructable in events with multiple sources of missing energy.  
For highly boosted tops, the finite angular resolution of the 
detector makes complete reconstruction of the system difficult, and angular distributions
in the top rest frame are no longer optimal observables.

Here we present a technique to measure the polarization of a boosted top in its hadronic
decay mode using the energy fraction distribution of a particular subjet.
% Our method does not rely on the assumption of the production and decay mechanism, and the reconstruction of the top rest frame.  Keeping in mind the challenge of tagging $b$-jets at high $p_T$, our new method works on subjets without such a requirement. 
%Such an analysis is not ideal.  It would be preferable not to have to
%make assumptions on the top production mechanism.  Also, if there are
%additional sources of missing energy one cannot fully reconstruct the
%top and must instead use techniques like $b$-tagging to infer its
%presence.  Tagging $b$-jets at high $p_T$, when they are nearly
%collinear with leptons, will be challenging.  Even more, by
%restricting oneself to semi-leptonic decays one pays a high
%statistical price due to the small leptonic branching fraction.
%Here we will present a technique that avoids all of these problems by
%using subjets to measure the polarization of the top in its hadronic
%decay mode, when it decays entirely into a single jet.  
This new method does not require high-$p_T$ $b$-tagging, which is known to be challenging.
We also do not require $W$ reconstruction inside the top jet.
Again, as we are considering hadronic tops, this technique measures top 
polarization using information from the top jet alone, independent of 
other objects in the event.  It does not involve the  reconstruction of top rest frame, or rely upon the measurement of 
missing momentum. 

While identification of boosted hadronic tops above the QCD background is challenging, many promising approaches have been developed~\cite{Barger:2006hm,Agashe:2006hk,Fitzpatrick:2007qr,Lillie:2007yh,Skiba:2007fw,Baur:2007ck,Frederix:2007gi,Baur:2008uv,Thaler:2008ju,Kaplan:2008ie,Almeida:2008yp,Almeida:2008tp,Bai:2008sk, cmsca, Brooijmans}. In this article we will assume that the boosted top candidates can be identified through one of these methods.

We will begin by motivating our choice of a subjet-based technique for
studying the substructure of a top-jet.  Then we will propose an algorithm useful
for measuring the top polarization and discuss its interpretation.
Finally, we will demonstrate the robustness of the algorithm by
testing it in different physics scenarios with data from different
parton shower models.

%%%%%%%%%%%%%%%%%%%%%%%%%%%%%%%%%%%%%%%%%%%%%%%%%%%%%%%%
\section{Looking Inside a Top Jet}
%%%%%%%%%%%%%%%%%%%%%%%%%%%%%%%%%%%%%%%%%%%%%%%%%%%%%%%%

Here we will discuss the different techniques used to study boosted
hadronic tops.  This will give us a chance to motivate our use of
subjets while outlining other possibilities.

In the past, two distinct approaches have been taken to analyze top
jets.  One approach uses jet shape
variables~\cite{Almeida:2008yp,Almeida:2008tp,Thaler:2008ju} to define
a function on the constituents of a top jet (in practice, the
constituents will be calorimeter cells), treating each constituent
independently.
%~\footnote{So that the result is a sum of contributions
%computed using each constituent four-momenta and the overall jet
%momenta.}  
The other approach~\cite{Thaler:2008ju,Kaplan:2008ie, cmsca, Brooijmans}
defines a function on the subjets formed by reclustering the
constituents of a larger jet.  Functions then depend upon the
constituent four-momenta only through the total four-momentum of the
subjet they are clustered into, rather than upon each constituent
four-momentum independently.

Each approach has both advantages and disadvantages.  Subjets can reduce
the effects of soft contamination\footnote{Contamination, radiation
clustered within the top jet that did not arise from the top decay,
can be the result of initial state radiation, multiple interactions,
or wide angle emissions from other parts of the event.} by summing
together constituents so that softer particles have a proportionally
smaller influence.  However, care must be taken because some
quantities one can form from subjets, such as invariant mass, can be 
extremely sensitive to calorimeter spacing and out-of-cone radiation. 
%Invariant mass is one
%such quantity, and although it might seem well suited for
%reconstructing the $W$ and thus identifying the $b$-jet, we avoid its
%use due to this sensitivity~\footnote{Even if one accepted a sensitive
%quantity, the smeared invariant masses and combinatorial effects make
%it difficult to identify the correct pair of $W$ decay products from
%among the subjets.}.  
Fortunately, as long as one avoids these
troublesome quantities a subjet-based analysis can be made fairly
robust.  For our algorithm below, we will only rely upon the relative
hardness and separation of the subjets, both quantities which are
fairly insensitive to additional soft radiation and detector effects.

Jet shape variables, because they treat each jet constituent independently,
are more amenable to higher order calculations than variables defined with
subjets.  However, these variables can become very sensitive to the
effects of contamination.  As an example, consider the {\it planar flow} jet shape
of~\cite{Almeida:2008yp}, which is equivalent (up to an overall
constant) to $\det S^\perp$ defined in~\cite{Thaler:2008ju}.  The
planar flow of a jet is defined as
\be
{\rm Pf}=\frac{4\lambda_1\lambda_2}{(\lambda_1+\lambda_2)^2}
\ee
where $\lambda_{1,2}$ are the two eigenvalues of the matrix
\be
I_w^{kl}=\sum_i w_i \frac{p_{i,k}}{w_i}\frac{p_{i,l}}{w_i}
\ee
where $w_i$ is the energy and $p_{i,k}$ the $k$th transverse momentum
component of the $i$th jet constituent.  This quantity essentially
decomposes the jet's radiation into two moments $\lambda_{1,2}$, 
similar to moments of inertia, so that if the jet is symmetric about
its center then ${\rm Pf}\approx 1$.  Planar flow is useful in
top-quark studies because top jets are relatively symmetric about
their center (corresponding to higher values of planar flow), while
QCD events are dominated by a single emission (corresponding to a
lower value of planar flow).  Unfortunately, planar flow weights each
constituent according to its transverse momenta relative to the jet
axis, so that as the radius of a jet is varied soft radiation towards
the edge of a jet begins to dominate and all events are skewed toward
higher ${\rm Pf}$\footnote{While soft effects can skew Pf toward higher values, it
remains difficult for QCD jets to be skewed all the way toward ${\rm Pf}\sim 1$.  Thus, while the distributions
of Pf are sensitive to soft effects, they can still be used as effective top/QCD discriminants.}.  To demonstrate this sensitivity and how it can 
be reduced through the use of subjets, we have included Fig.~\ref{fig:pf_comp}, 
showing the calculation of planar flow at matrix element level, after 
showering, and after reclustering using subjets.  The subjets are formed 
using the procedure described in Section~\ref{sec:examples} using $R=0.2$ cones.  
Here one can see the large corrections to the matrix element results that are 
attributable to soft radiation.
Of course, one can mitigate this effect by using smaller cones (the authors of~\cite{Almeida:2008yp}
used $R=0.4$), as the amount of diffuse soft radiation clustered into the top jet goes roughly as $R^2$, yet even in this
regime the effect of soft contamination can still be significant, especially near ${\rm Pf}\approx 0$.

\FIGURE[t]{
\includegraphics[scale=.35]{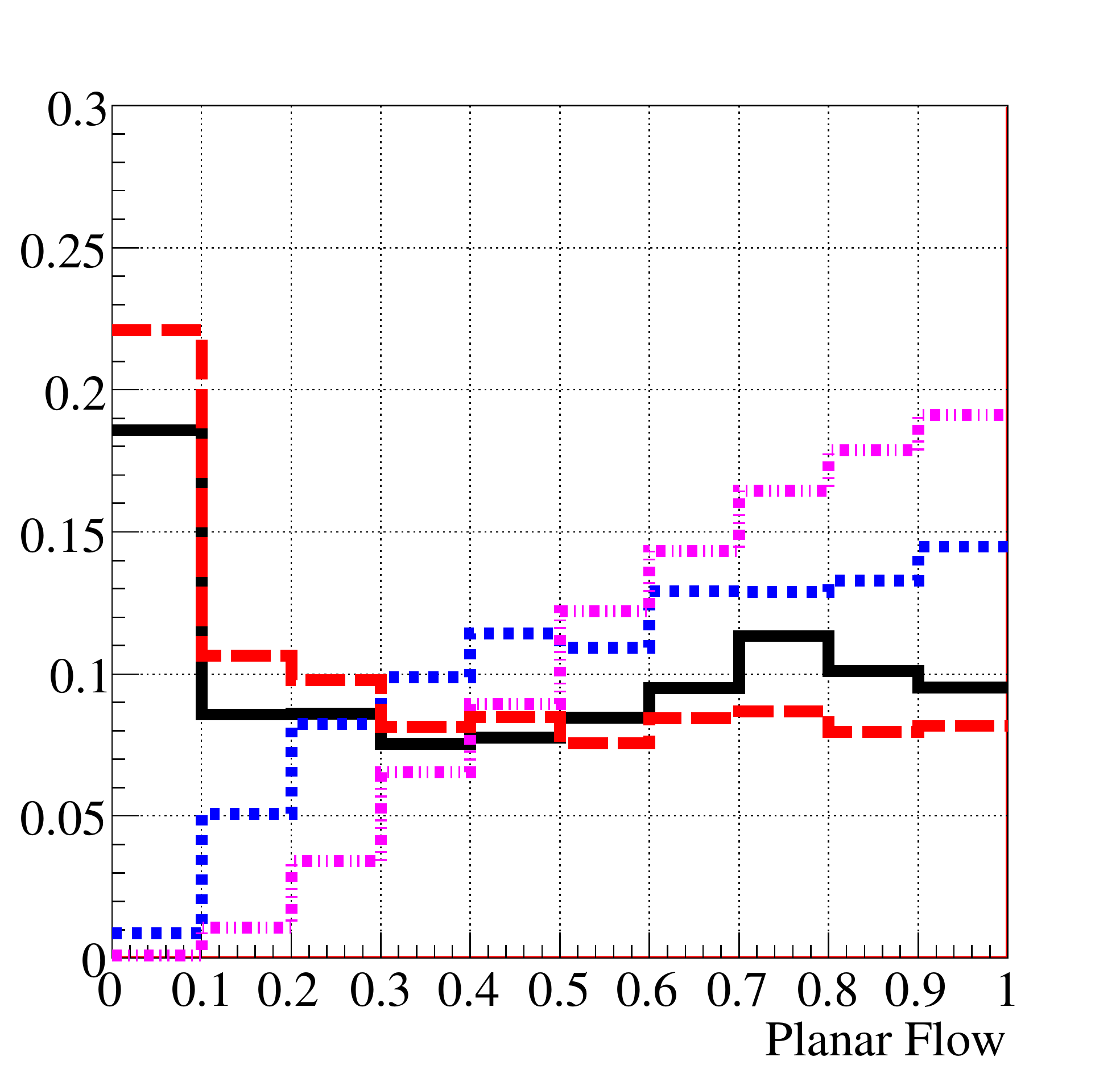} 
\includegraphics[scale=.35]{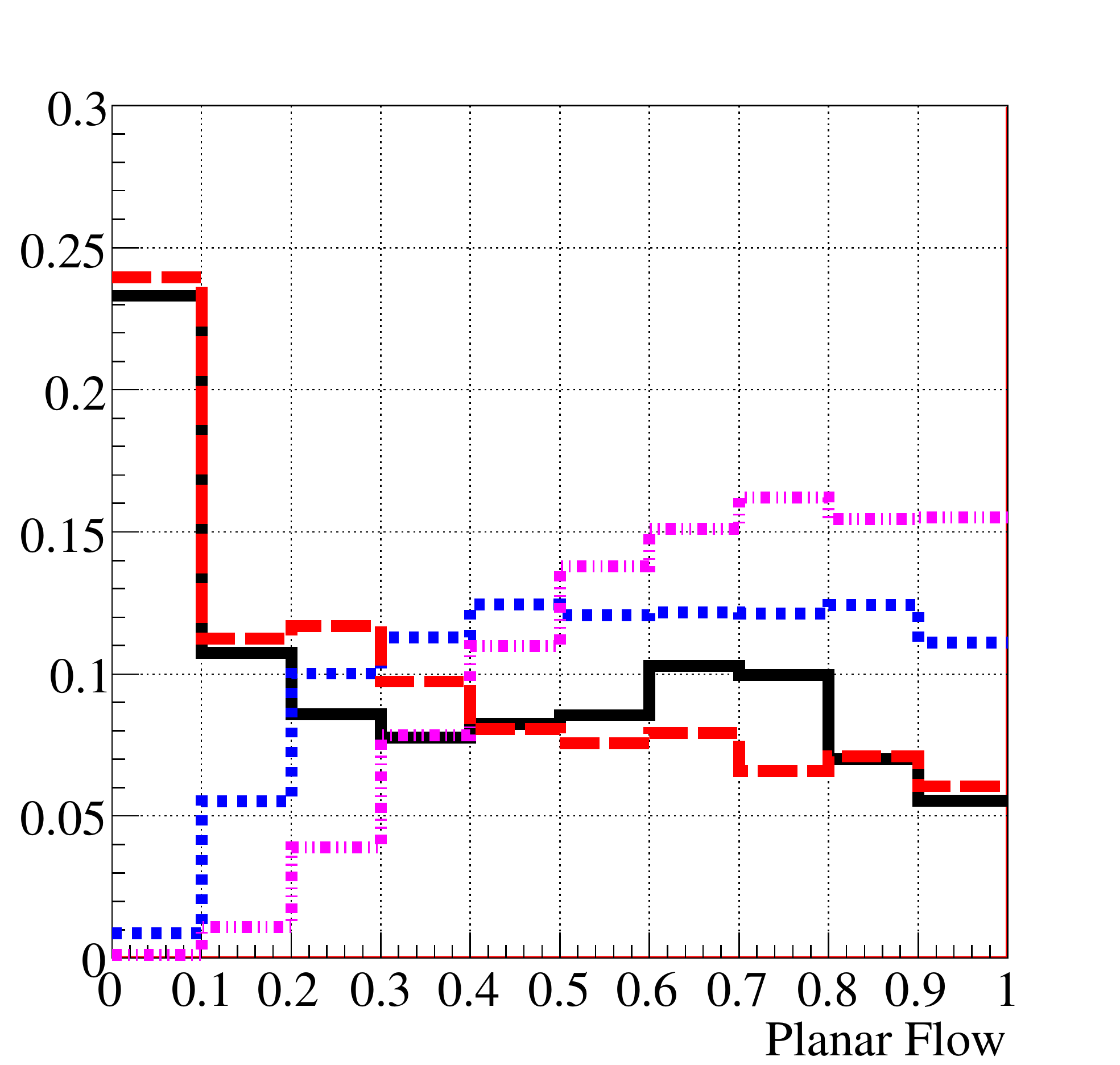} 
\caption{
\label{fig:pf_comp}
Comparison of planar flow for left-handed tops (left figure) and right-handed tops (right figure).  For each set of samples we compute the planar flow using the three partonic decay products of the top (black, solid), the constituents of the showered jet using a small  $R=0.4$ cone (blue, dotted), the constituents of the showered jet using a larger  $R=1.0$ cone (magenta, dot-dashed), and the three $R_{\rm sub}=0.2$ subjets formed from the $R=0.4$ top-jet constituents (red, dashed).  These events 
are taken from the decay of a $3~{\rm TeV}$ $Z'$ into two tops, clustered using the anti-$k_T$ algorithm, where we have required the top jet's mass satisfy  $140~{\rm GeV} < m_J < 210~{\rm GeV}$.
Note that the peak of the red/black distributions near ${\rm Pf}\sim 0$ can be ascribed to configurations where two of the partons become collinear, or where one parton becomes particularly soft.
}  }

To be sure, jet shape variables (including planar flow) are very
useful and will likely play a role in boosted top chirality
measurements.  A simple counting exercise shows that after
requiring the reconstruction of the $W$ mass, top four-vector, and
allowing for axial symmetry, there are still three remaining degrees of freedom
encoded in the matrix element that can be mapped out by jet shape
variables.  However, to simplify the discussion and avoid
complicated issues of contamination and higher-order corrections we
will use the rest of the paper to focus on quantities calculated using
subjets.

%%%%%%%%%%%%%%%%%%%%%%%%%%%%%%%%%%%%%%%%%%%%%%%%%%%%%%%%
\section{Top Polarimetry With Subjets}
%%%%%%%%%%%%%%%%%%%%%%%%%%%%%%%%%%%%%%%%%%%%%%%%%%%%%%%%

We will now explore methods for using subjets to measure the
polarization of a collimated hadronic top.  In what follows, we will
assume we are working with jets tagged as tops, as discussed
in~\cite{Thaler:2008ju,Almeida:2008yp,Almeida:2008tp,Kaplan:2008ie,Ellis:2009su},
and subsequently decomposed into three subjets (a prescription for
such a decomposition is given later).

%We begin with a collimated top jet, which we assume has already been tagged as coming from a top.
%We outline an algorithm which selects a subjet which serves as a good polarimeter for boosted hadronic
%tops.  

%%%%%%%%%%%%%%%%%%%%%%%%%%%%%%%%%%%%%%%%%%%%%%%%%%%%%%%%
\subsection{Choosing a Polarimeter} 
%%%%%%%%%%%%%%%%%%%%%%%%%%%%%%%%%%%%%%%%%%%%%%%%%%%%%%%%

One observable sensitive to the polarization of the top is the
distribution of energy among the its three decay products in the lab
frame.  In the collinear limit, the lab-frame energy fraction of the
$i$th subjet, $z_i ={E}_i/ {E}_{\rm{top}}$, depends only on the energy
and angular distributions in the top rest frame, and can serve as a
robust variable to measure polarization.  While energy fractions are
not Lorentz invariant for finite top mass, and in particular are not
invariant under longitudinal boosts, frame dependence enters only at
order $m_t/E_t $, and therefore, for highly boosted top quarks, energy
fraction variables become fixed, stable quantities~\footnote{Depending on the boost of the top quark, it might also be
desirable to consider subjet $p_T$ fractions, as the $m_t/E_t $ corrections to the collinear limit
differ for energy and $p_T$ fractions.}. The question then becomes how to
select the subjet to be used as a polarimeter.

The most obvious candidate for the job is the $b$-jet~\cite{Shelton:2008nq,Perelstein:2008zt}, identified either directly through $b$-tagging or indirectly by first finding the light quarks from the $W$.  
However, the identification of the $b$ and 
$W$  poses some experimental difficulties.
Even in isolation, the efficiency of $b$-tagging drops by a factor of $2$--$3$ at high $p_T$ while
light quark rejection is degraded by roughly a factor of 3 \cite{Lillie:2007yh,Baur:2007ck,March:851053,Gonz‡lezdelaHoz:814346,Lehmacher:2008hs}.
When the $b$-jet is situated within a collimated top jet, the additional tracks from the neighboring light quark subjets
present added complications for $b$-tagging algorithms.

Another possible method of identifying the $b$-jet is to do so
indirectly, by finding the $W$.  One approach to identifying the $W $
is to look for two jets with an invariant mass within the $W $ mass
window.  However, the subjet invariant mass distributions are
distorted both by contamination from soft radiation and by imperfect
subjet reconstruction, as well as by the finite size of the
calorimeter.  The invariant mass $m_{ij}$ of two nearby subjets is
approximately proportional to their separation in $R$, and for subjets
whose centers are separated by $\Delta R_{ij}\lesssim0.5$, the
uncertainty associated with the calorimeter granularity $\delta R\sim
0.1$ can be significant.  Distinguishing the correct $W$ subjet pair
from amongst the three choices, all of which are typically within a
factor of two of each other, then becomes difficult.

Another possible strategy to identify the $b$-jet is to look for hard
splittings within the top jet.  As discussed in~\cite{Thaler:2008ju},
the energy sharing of a parton branching $A\to B C$,
\be
z (A\to BC) \equiv \min (E_B, E_C)/E_A,
\ee
discriminates between hard splittings from decays, $z \sim 0.5$, and
soft splittings more characteristic of QCD, $z\sim 0$.  If the $W$
decay products were well-separated from the $b$-jet, one could
identify the $b$ by unwinding the clustering of the top jet until
there were two subjets and tagging the $b$ as the one with smaller $z$
(so the $W$ subjet would be the one with a harder splitting).
Unfortunately, because the $W$ has a mass on the same order as that of
the top, the $W$ decay products are not well-separated from the $b$,
so upon unwinding the top jet by one step one often finds that the
$b$-jet has been clustered with a lighter jet from the $W$ decay.

We propose here an alternate subjet selection algorithm, based on $k_T
$ distances between subjets, which does not require either $b$ or $W $
identification.  While the algorithm is conceptually less
straightforward than those based on attempting to identify specific
partons within the top jet, it yields a distinct separation between
chiralities and is robust under showering and detector effects.
Consider the $k_T$ distance measure between two four-momenta $i$ and
$j$,
\be
d_{ij} =\min (p_{Ti} ^ 2, p_{Tj} ^ 2) R_{ij} ^ 2,
\ee
where $R_{ij}^ 2 = (\eta_i-\eta_j) ^ 2+ (\phi_i-\phi_j) ^ 2$.  Of the
three $d_{ij}$ one can form from the three top subjets, consider the
smallest.  Our top polarimeter is the energy fraction $z_K$ of the
{\it harder} jet $j_K$ in the minimum $k_T$ distance pair.  We plot
the distribution of this variable at parton level for different
chiralities in Fig.~\ref{fig:ktz}.  The variable shows a clear
distinction between right- and left-handed top quarks, with
right-handed tops peaked at smaller values of $z_K $, and left-handed
tops preferring larger values of $ z_K $.
%%%%%%%%%%%%%%%%%%%%%%%%%%%%%%%%%%%%%%%%%%%%%%%%%%%%%%%% 
%% Figure: E frac of kT-selected jet
%%%%%%%%%%%%%%%%%%%%%%%%%%%%%%%%%%%%%%%%%%%%%%%%%%%%%%%% 
\FIGURE[t]{
\includegraphics[scale=.4]{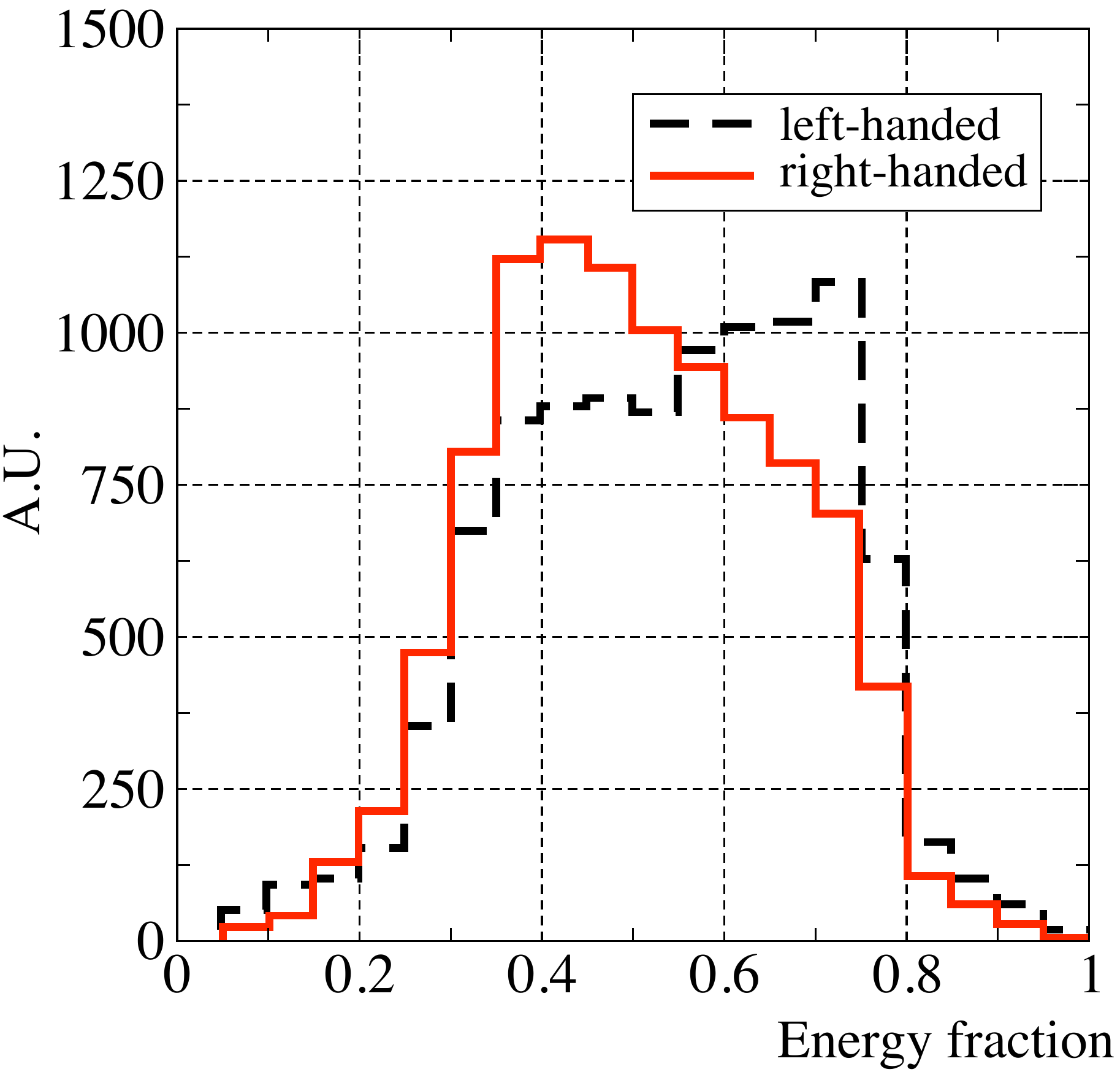} 
\caption{
\label{fig:ktz}
Energy fraction $z_K$ of the parton selected by the $k_T $-based algorithm for different top polarizations.  The events shown
here correspond to tops produced from a $3 \tev$ resonance.}  }
%%%%%%%%%%%%%%%%%%%%%%%%%%%%%%%%%%%%%%%%%%%%%%%%%%%%%%%%

%%%%%%%%%%%%%%%%%%%%%%%
\subsection{Operation of the algorithm}
%%%%%%%%%%%%%%%%%%%%%%%

The success of the jet $j_K$ selected by this algorithm as a polarimeter depends on
multiple aspects of the angular and energy distributions of daughter partons in
polarized top decay, which for reference are reviewed in the Appendix.
In order to explain the success of our polarimeter, 
we first consider how the algorithm functions at parton level.  The identities
of the partons picked out by the algorithm differ between right- and left-handed tops.
In Fig.~\ref{fig:Efracbreakdown} we break down the contributions to the variable
$z_K$ by parton identity.
%%%%%%%%%%%%%%%%%%%%%%%%%%%%%%%%%%%%%%%%%%%%%%%%%%%%%%%%
%% Figure:  E frac breakdown by daughter identity
%%%%%%%%%%%%%%%%%%%%%%%%%%%%%%%%%%%%%%%%%%%%%%%%%%%%%%%%
\FIGURE[t]{
\includegraphics[scale=.4]{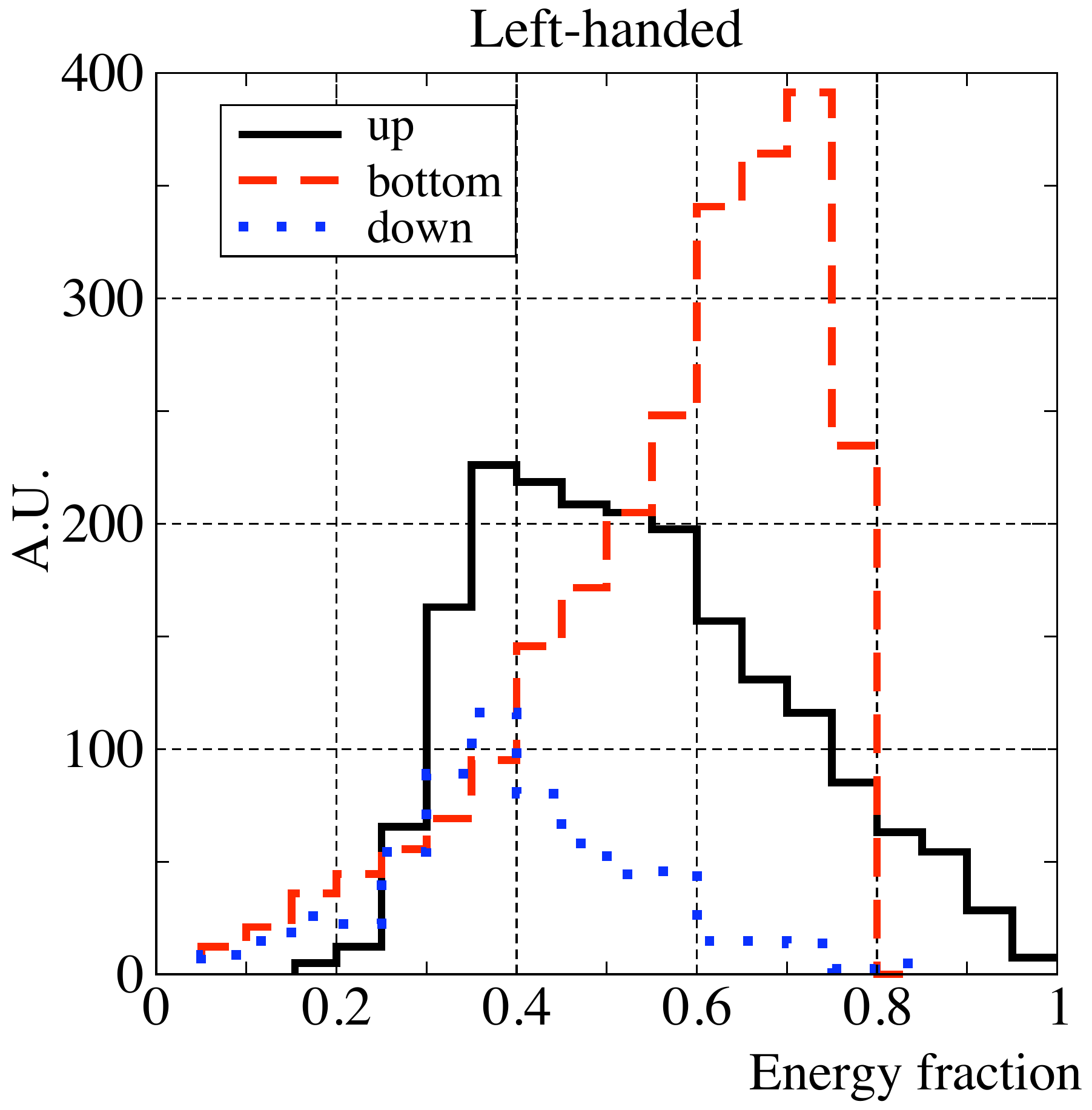}
\includegraphics[scale=.4]{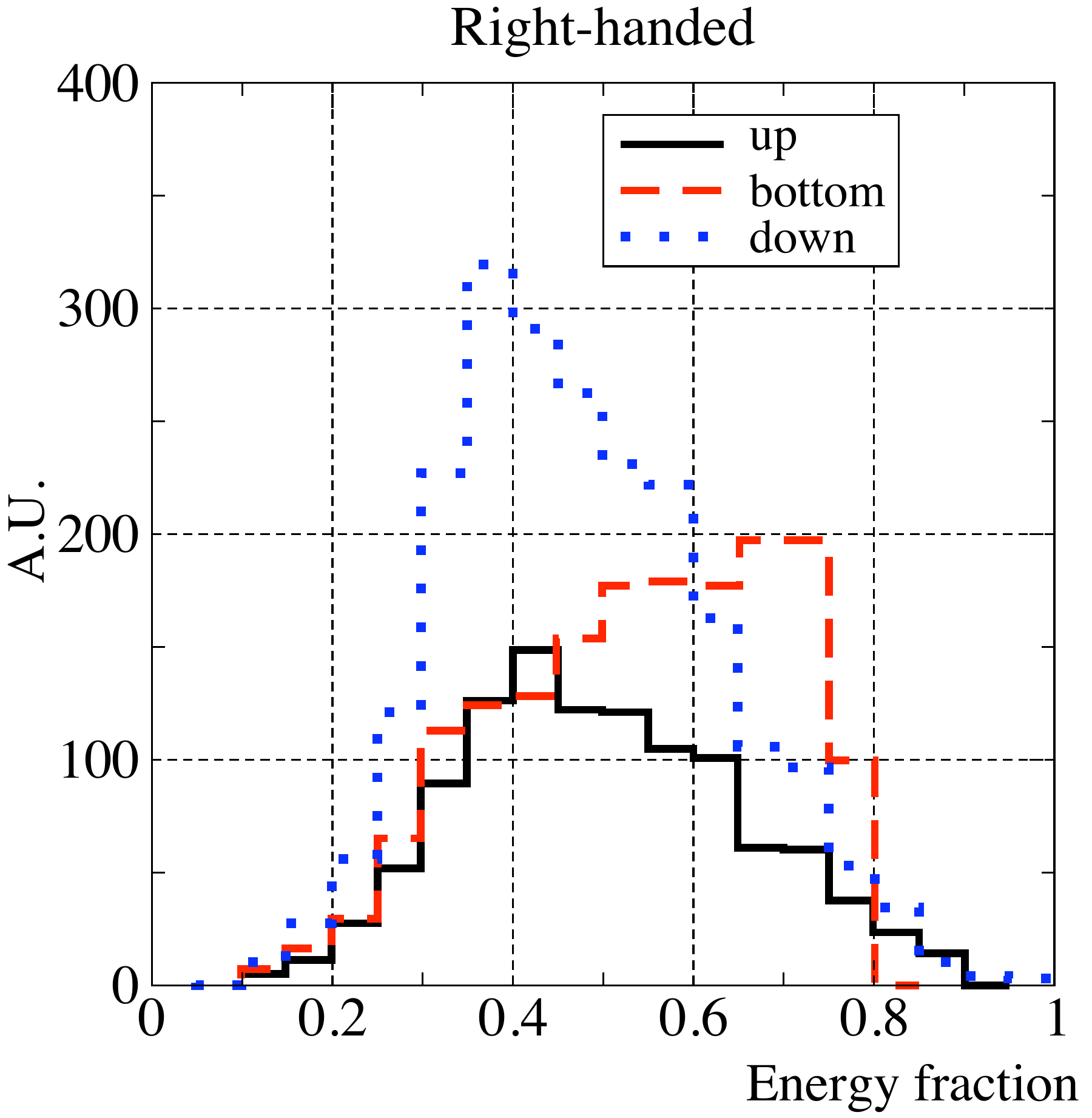}
\caption{\label{fig:Efracbreakdown}Energy fraction of the parton selected by our algorithm,
broken down by parton identity.  The events shown here
correspond to tops produced from a $3 \tev$ resonance.}
}
%%%%%%%%%%%%%%%%%%%%%%%%%%%%%%%%%%%%%%%%%%%%%%%%%%%%%%%%

The anti-down quark is maximally correlated with the top spin, and
thus for left-handed tops the $d$ tends to be soft.  For left-handed tops
the minimum-$k_T$
pair therefore tends to involve the $d$, and in such pairs the other
parton ($b$ or $u$) is the harder of the two.  The algorithm therefore
picks out first $b$ quarks, which take a larger fraction of the top
energy, and secondarily $u$'s, with $d$ quarks a distant third.

For right-handed tops, where the top energy is shared more equitably
among the daughter partons, angular correlations play a more central
role.  The $d$-quark is now both more central and harder than
predicted by pure phase space (due, again, to its maximal correlation
with the top spin).  Therefore in order to reconstruct the necessary
invariant masses, the $\Delta R$ separation between the $d $ and the
$u$ and $b$ quarks must be smaller than for pure phase space, and the
minimum $k_T$ pair then tends to involve the $d$.  For right-handed
tops, the algorithm thus dominantly selects the $d$-quark, as can be
seen in Fig.~\ref{fig:bzbreakdown}.  While the $d$ is preferentially
emitted along the top direction of motion, its energy fraction
distribution nonetheless falls off at high energies, as the lab-frame
$d$-quark energy fraction depends on the energy of the $d$-quark in
the top rest frame as well as the angle of emission.  The contribution
of the $b$-quark to the variable $z_K $ comes mostly from hard $b$'s
recoiling against soft transverse $W$'s.

At high parton energy fraction $z_K$, the algorithm dominantly selects
the hardest parton: $b$ and $u$ for left-handed tops, $b$ and to a
lesser extent $u, d $ for right-handed tops.  At intermediate energy
fractions, the origin of the $u$ and $d $ partons from a common $W$
comes to dominate.  The $k_T$ distance between the $u$ and the $d$
is bounded from above, as the $u$ and the $d$ must reconstruct the
$W $.  In events without hierarchical energy distributions, the
minimum $k_T$ distance thus tends to be between the decay products of
the $W$.  Therefore, at intermediate energy fractions, the parton
selected by the algorithm is predominantly the $d$ (for right-handed
tops) or the $u$ (for left-handed tops).  This can be seen already in
Fig.~\ref{fig:Efracbreakdown}, and is further demonstrated in
Fig.~\ref{fig:bzbreakdown}.

Finally, we note that all of these arguments are based upon the 
assumption that one can go from the collider coordinate system to one 
oriented around the top direction of motion without significant effects.  
This assumption does not hold exactly, because the detector geometry is not invariant under rotations around the
axis defined by the top direction of motion, and because the $k_T$
algorithm used to select the subjet $j_K$ makes reference to the
collider coordinate system through the definition of transverse
momentum.  Therefore, events which differ from each other only by a rotation around
the top axis of motion appear differently both in the detector and in
the subjet selection algorithm.  Interference terms between right- and
left-handed tops generically then do not completely cancel.  However, as the magnitude of the
interference contribution is determined by the components of the
parton momenta transverse to the top momentum, these effects are of
order $m_t/E_t$, a subleading effect for large boosts.

%%%%%%%%%%%%%%%%%%%%%%%%%%%%%%%%%%%%%%%%%%%%%%%%%%%%%%%%
%% Figure:  min kt breakdown
%%%%%%%%%%%%%%%%%%%%%%%%%%%%%%%%%%%%%%%%%%%%%%%%%%%%%%%%
\FIGURE[t]{
\includegraphics[scale=0.35]{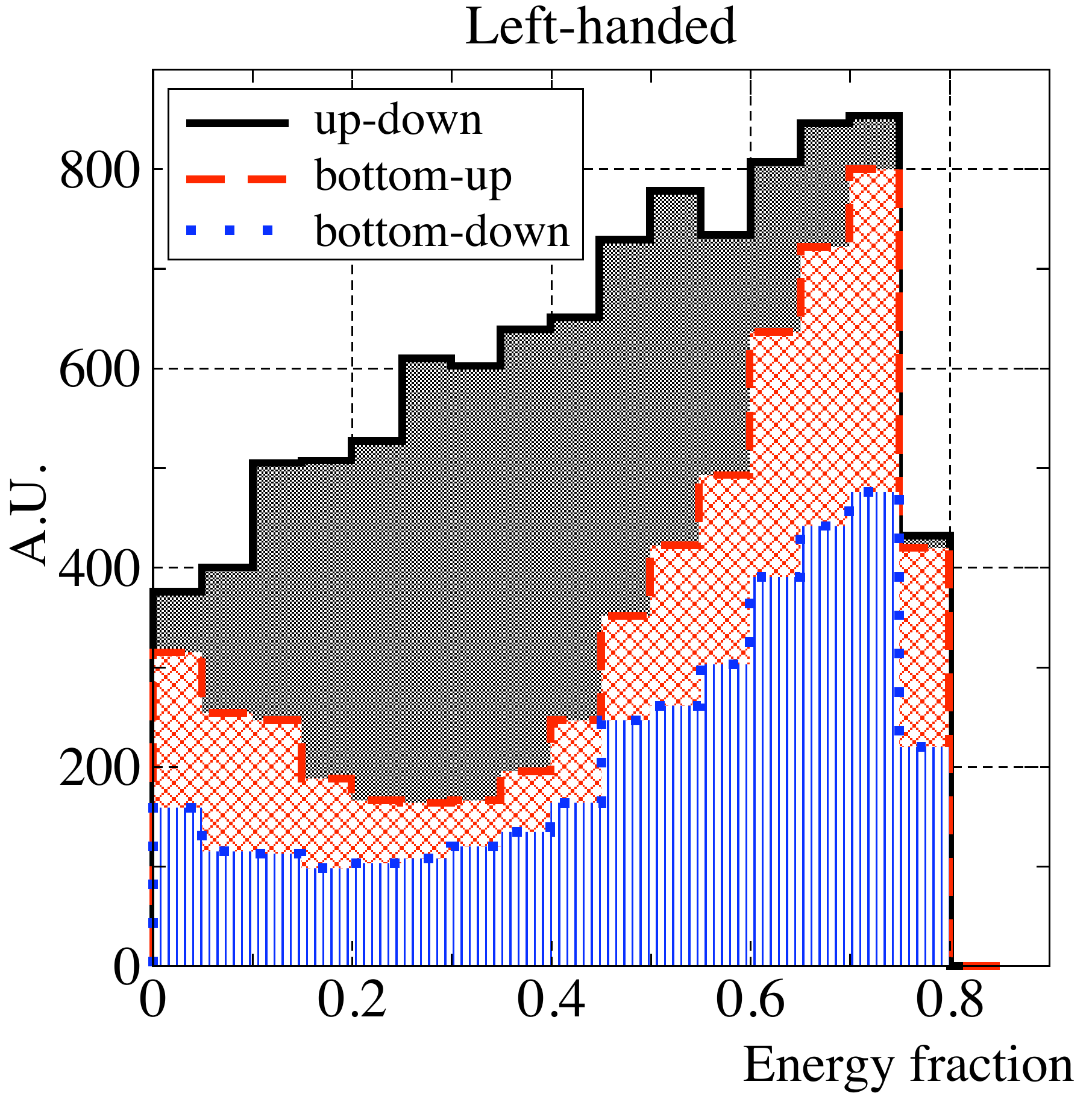}
\includegraphics[scale=0.35]{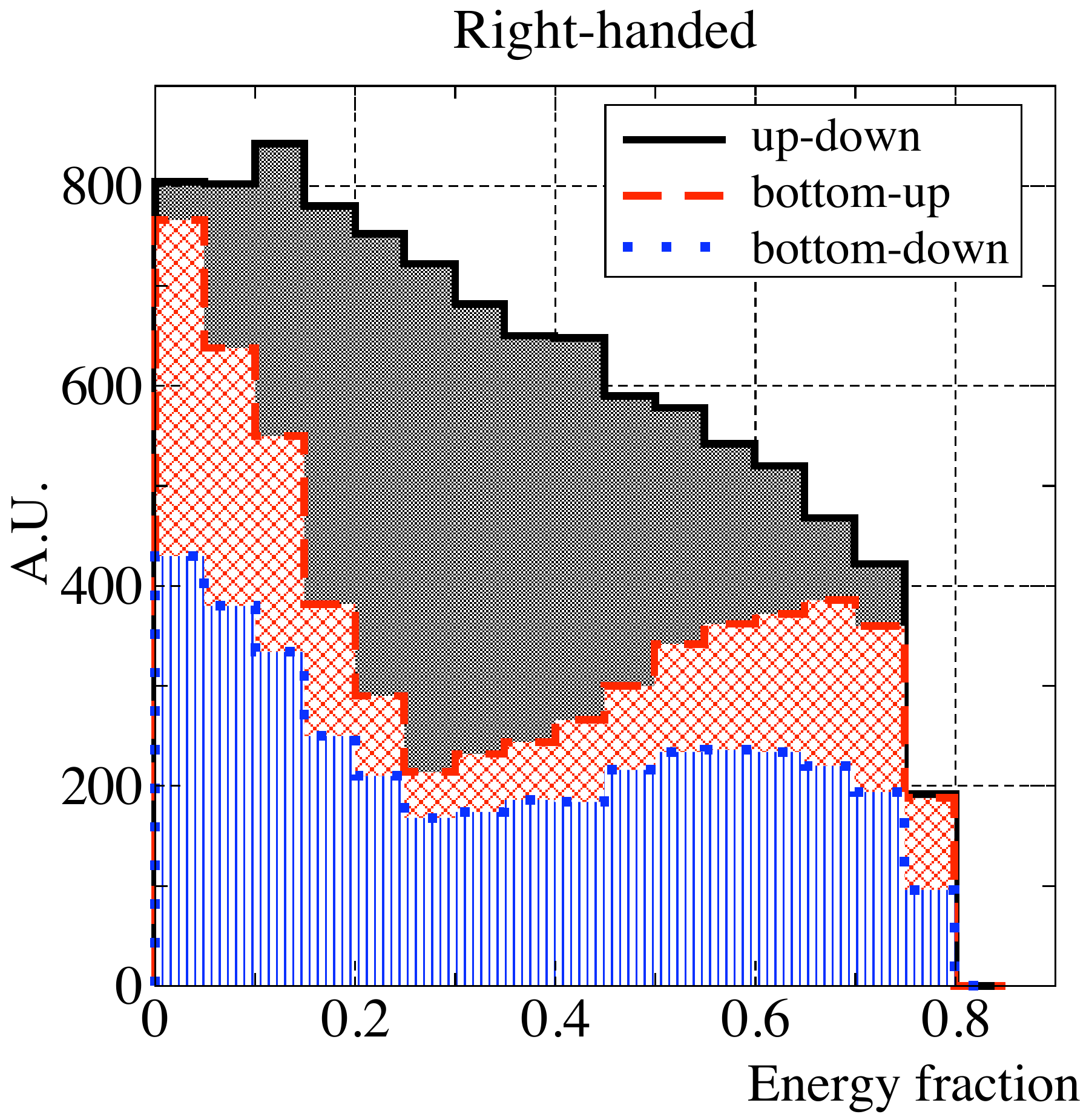}
\caption{Energy fraction of the $b$ quark, broken down according to which two 
partons in the event belong to the pair with minimum $k_T$ for left-handed tops (left side) and right-handed tops (right side).  Note that the contents of the plots are stacked.
At high energy fractions, the $W$ and its decay products are soft, and the
minimum $k_T$ pair tends to involve one of the $W $ decay products together with the $b $.
At intermediate energy fractions, the minimum $k_T$ pair tends to be the $W$ 
decay products.  At small energy fractions, the $b$ begins to appear as the softer of 
the two partons in the minimum $k_T$ pair.  The effect is more pronounced for
right-handed tops, which have a less hierarchical distribution of energy among
the three daughter partons.
The events shown correspond to tops produced from a $3 \tev$ resonance.
\label{fig:bzbreakdown}
}
 }
%%%%%%%%%%%%%%%%%%%%%%%%%%%%%%%%%%%%%%%%%%%%%%%%%%%%%%%%

%%%%%%%%%%%%%%%%%%%%%%%%%%%%%%%%
\subsection{Implementation}
%%%%%%%%%%%%%%%%%%%%%%%%%%%%%%%%

To implement this algorithm in practice one must have a technique for 
finding three subjets within the top jet.  The exact procedure one 
uses to identify the subjets is not important, but for concreteness we 
detail the method used in our study.  One advantage of our 
subjet-finding technique is that it is easy to implement within 
the FastJet~\cite{Cacciari:Fastjet} framework, already used by many 
studies for top tagging.

The procedure is as follows:
\bi
\item Cluster the event with a reasonably sized cone ($\Delta R
\gtrsim 0.7$) and select a top candidate.  

\item Take all the cells clustered into the top candidate and
recluster them using a smaller cone ($\Delta R \approx 0.2$).

\item Demand that there are at least three subjets, each with a
substantial amount of the jet's energy $\gtrsim 1-2\%$.  If there are
not, split the harder subjet by unwinding~\cite{Thaler:2008ju} it one
step using the $k_T$ algorithm, and use the two resulting daughters
along with the second hardest subjet from before.

\item Now use the four-momenta of the three subjets to find the pair
with the smallest $k_T$ distance measure, and compute the ratio of the
energy of the more energetic jet in this minimum-$k_T $ pair to the
energy of the entire top jet.
\ei
Results using this procedure are shown in the next section.

%%%%%%%%%%%%%%%%%%%%%%%%%%%%%%%%%%%%%%%%%%%%%%%%%%%%%%%%
\section{Examples}
\label{sec:examples}
%%%%%%%%%%%%%%%%%%%%%%%%%%%%%%%%%%%%%%%%%%%%%%%%%%%%%%%%

We will now apply the subjet-based technique developed in the previous section to some
realistic examples.  Our goal is to show that the technique works for
fully showered events clustered with finite calorimeter cells using a
variety of parton shower and hadronization algorithms.  It
is important to note that we do not consider the shape of background
QCD distributions, nor do we consider any shaping effects that might
arise from the effects of top tagging.  A more complete experimental
study would include these effects, but due to the high discriminating
power of top tagging algorithms (not to mention other aspects of the
event that could be used to remove background) and their relatively
high efficiency, we do not expect these effects to be significant.

In what follows, our analysis is performed on events generated at
matrix element level using \texttt{MadGraph
4.4.17}~\cite{Maltoni:2002qb} for physics at the LHC scale ($14\tev$).
Subsequent showering and hadronization is performed using
\texttt{Pythia 6.4.21}~\cite{Sjostrand:2006za} and \texttt{Herwig++
2.3.2}~\cite{Bahr:2008pv}.  When using Pythia, we consider parton
showers generated using both virtuality (labeled as $Q^2$) and $p_T$
ordered showers.  Visible final state particles are grouped into
$0.1\times 0.1$ calorimeter cells before being clustered into $R=0.7$
jets using the anti-$k_T$~\cite{Cacciari:2008gp} algorithm.  To ensure
that the top decayed into visible products (and that no significant
radiation was lost outside the cone) we demand that the jet mass
exceed $170~{\rm GeV}$.  We form subjets by running the anti-$k_T$ algorithm with
$R=0.2$ on the constituents of the top jet, requiring that the third
most energetic subjet carry at least $1\%$ of the top jet energy, and
splitting the hardest subjet if this condition is not satisfied.

%%%%%%%%%%%%%%%%%%%%%%%%%%%%%%%%%%%%%%%%%%%%%%%%%%%%%%%%
\subsection{Tops from a resonance}
%%%%%%%%%%%%%%%%%%%%%%%%%%%%%%%%%%%%%%%%%%%%%%%%%%%%%%%%

We begin by studying a colored octet vector $G'$ with a chiral coupling to 
the top quark.  This model was chosen for simplicity, but it captures 
the main features of well-motivated scenarios like KK-gluon production.  
The process under consideration is 
\be
gg\rightarrow G'\rightarrow t\bar{t}
\ee
where $m_{G'}=3~\rm{TeV}$.  The results are shown in Fig.~\ref{fig:simzp}. 
\FIGURE[t]{
\includegraphics[scale=0.225]{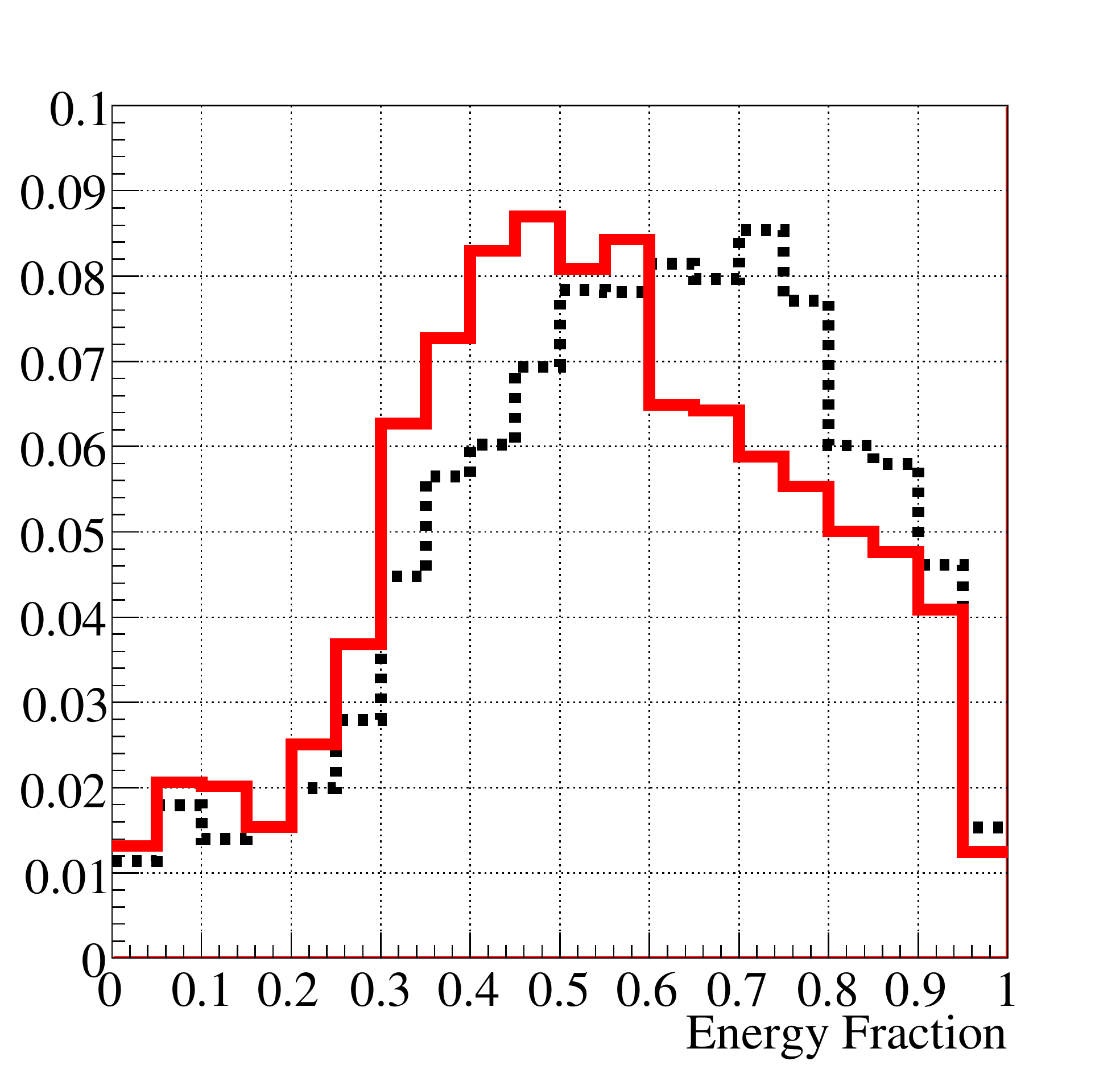}
\includegraphics[scale=0.225]{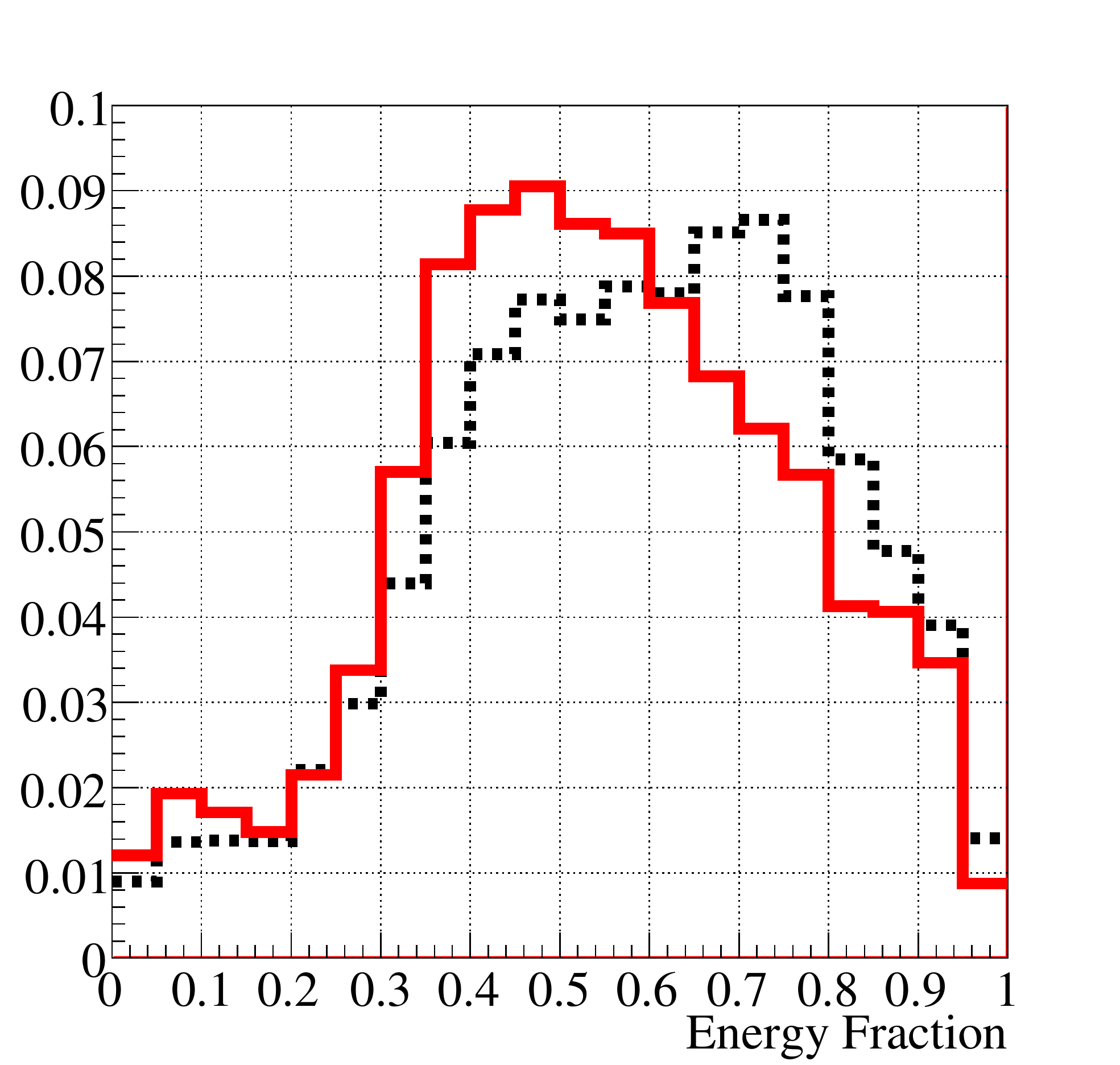}
\includegraphics[scale=0.225]{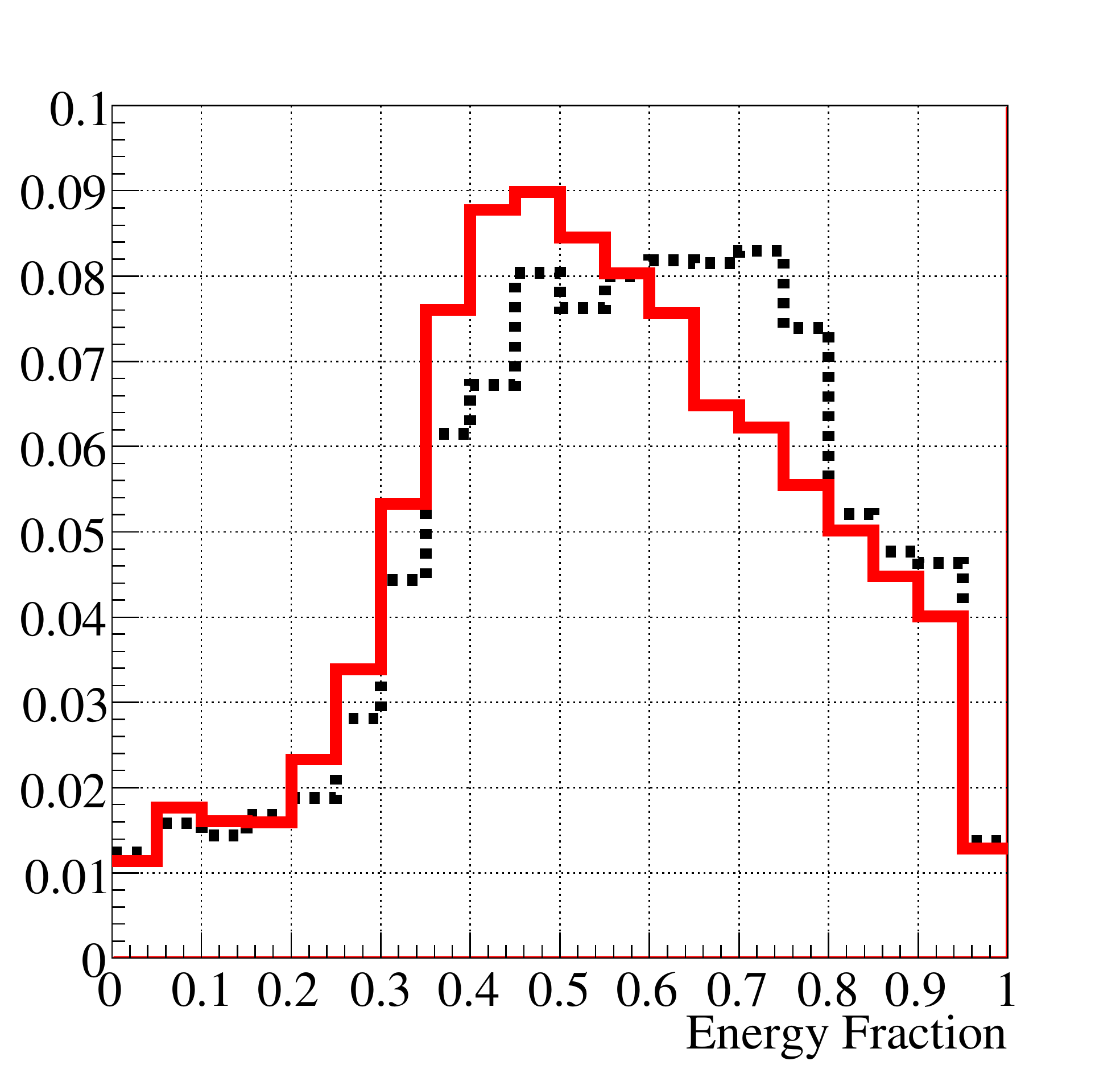}
\caption{Energy fraction of the selected jet for results from (left to right) Herwig++, Pythia-6 ($Q^2$) and Pythia-6 ($p_T$).  For each plot the solid red and dotted black lines come from right- and left-handed tops, respectively.
\label{fig:simzp}
} }
One can see from these distributions that the characteristic shapes from 
matrix element level are unchanged after parton showering and hadronization, 
demonstrating the robustness of our subjet selection technique.
%supporting our claim of a robust technique.  

%%%%%%%%%%%%%%%%%%%%%%%%%%%%%%%%%%%%%%%%%%%%%%%%%%%%%%%%
\subsection{Tops from cascade decays}
%%%%%%%%%%%%%%%%%%%%%%%%%%%%%%%%%%%%%%%%%%%%%%%%%%%%%%%%

\FIGURE[t]{
\includegraphics[scale=0.225]{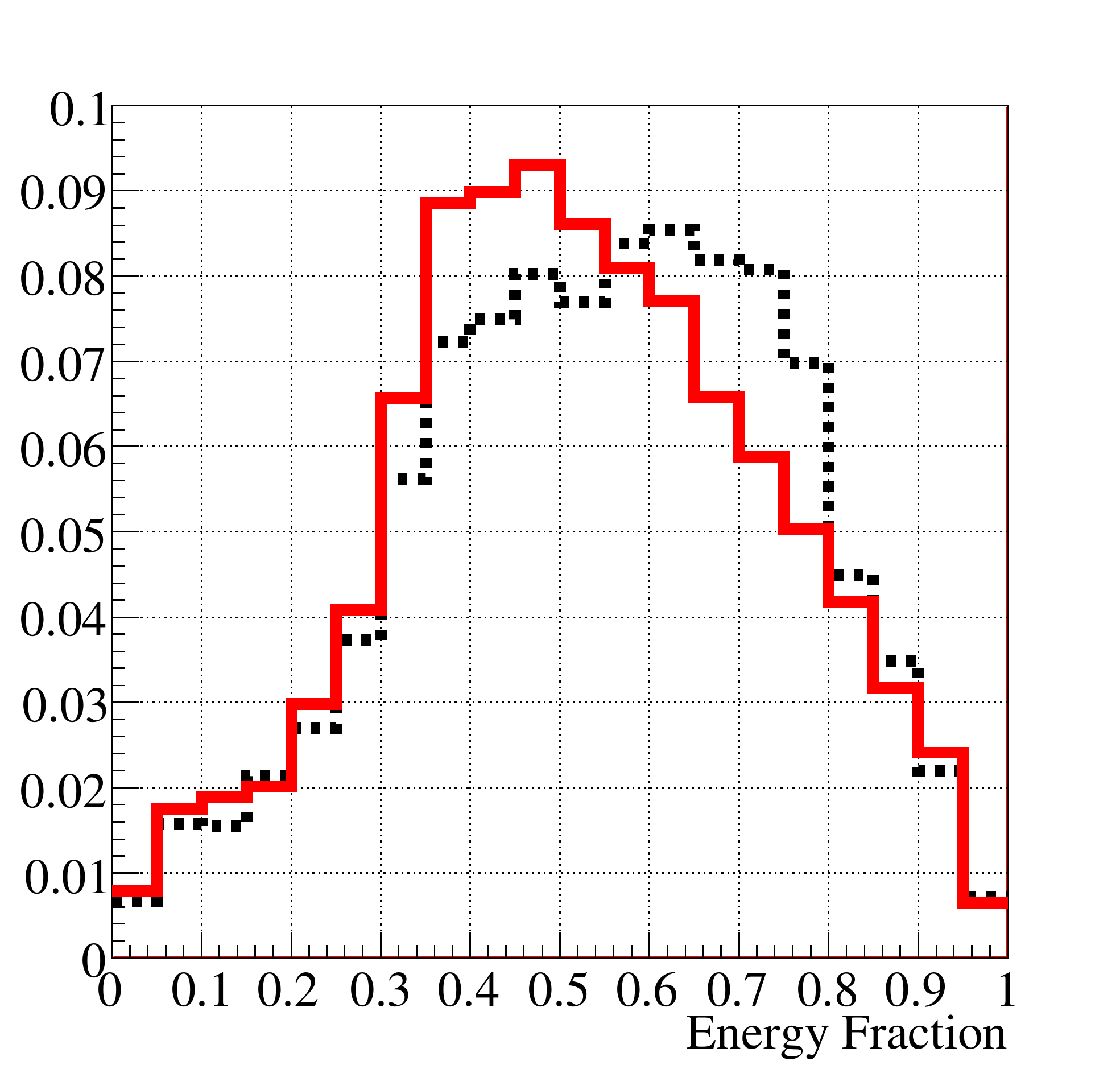}
\includegraphics[scale=0.225]{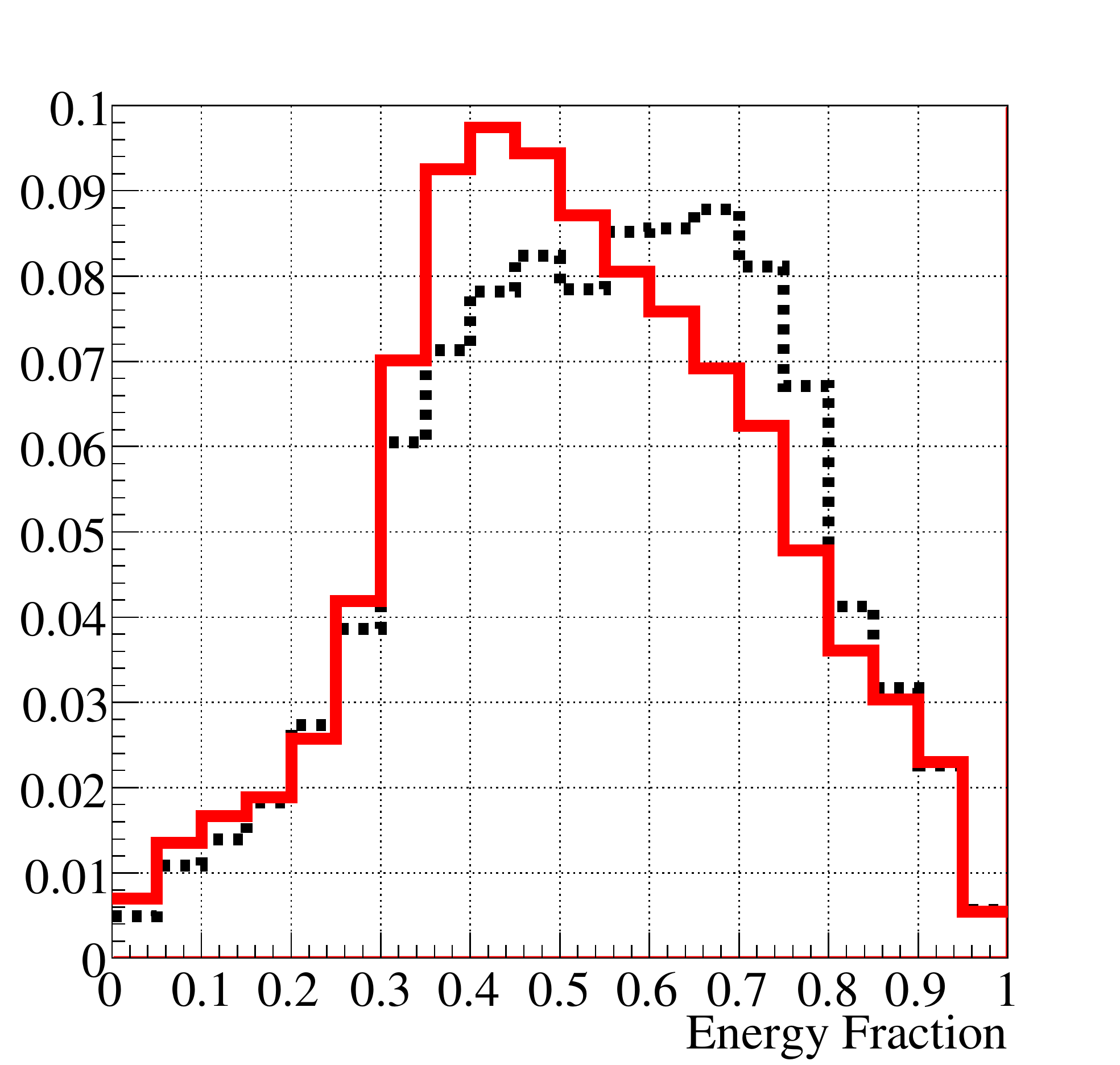}
\includegraphics[scale=0.225]{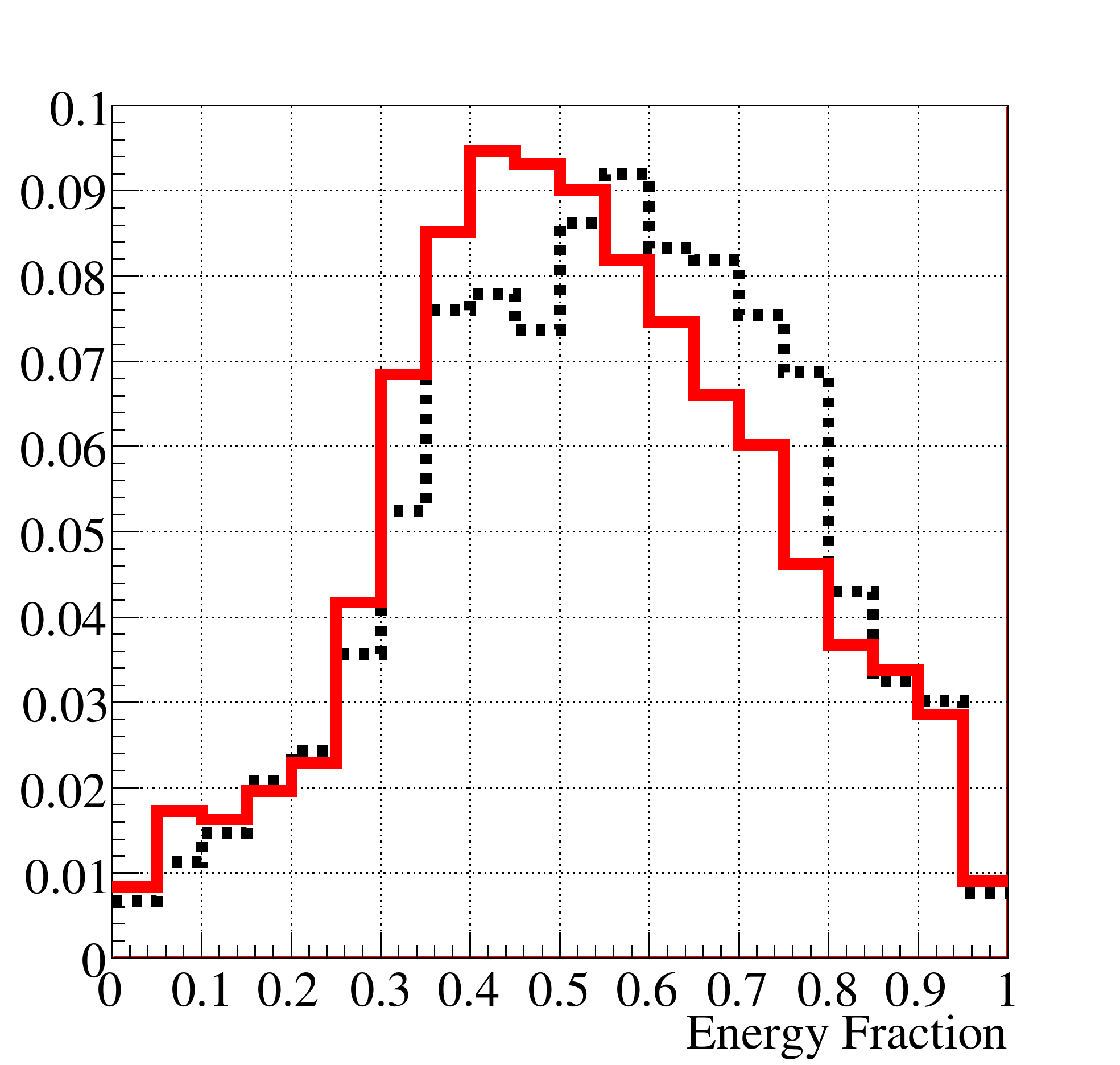}
\caption {Energy fraction of the selected jet for results from (left to right) Herwig++, Pythia-6 ($Q^2$) and Pythia-6 ($p_T$).  For each plot the solid red and dotted black lines come from right- and left-handed tops, respectively.
\label{fig:simtp}
}
 }
%\FIGURE[h!]{
%\includegraphics[scale=0.6]{top_partner2}
%\caption{$T'$ production and decay into top plus missing energy.
%\label{fig:cascade}
%}
%}

Cascade decays of an on-shell top partner (such as a stop squark or a
$T'$) to a top plus missing energy are a standard signal of a broad
class of well-motivated models.  In the presence of multiple sources
of missing energy, there is no longer enough information to solve for
the rest frame of a leptonically-decaying top quark.  Hadronic top
quarks, which can be reconstructed independently of the other
particles in the event, become a more useful source of information.

For tops produced from an un-reconstructed parent, the observable
polarization signal depends on the masses of the new physics particles
through the relation of the unknown parent rest frame to the lab
frame, as well as through the vertex kinematics \cite{Shelton:2008nq}.
The lack of information about the parent rest frame reduces the
observable polarization signal for tops coming from a cascade decay
compared to the signal from a resonance.  Nonetheless, observable
signals are still possible as long as the boost of the top from its 
parent is dominant, allowing the chiral structure of the top
production vertices to be probed.

We consider a model for production of two top
partner $T'$ particles decaying into tops and sources of missing
energy (labeled $A^0$).  Our choice of spectrum has $m_{T'}=2\tev$ and
$m_{A^0}=100~\rm{GeV}$.  The results of the analysis performed on the
model are seen in Fig.~\ref{fig:simtp}.  As for the $G'$,
the distributions agree well with the parton level results and have
the same qualitative shapes regardless of the generator used.

%%%%%%%%%%%%%%%%%%%%%%%%%%%%%%%%%%%%%%%%%%%%%%%%%%%%%%%%
\section{Conclusions}
%%%%%%%%%%%%%%%%%%%%%%%%%%%%%%%%%%%%%%%%%%%%%%%%%%%%%%%%

We have proposed an analysis tool useful in determining the chiral
structure of the top quark's coupling to new physics.  Our method uses
subjet-based techniques to probe scenarios where a highly boosted top  
decays hadronically.  This tool requires no assumptions to be made about the
production mechanism of the top or about the origin of missing
energy in the event, and does not rely upon $b$-tagging or $W$
reconstruction.

By testing our method on Monte Carlo data from multiple generators
using different new physics scenarios we have indicated its
robustness against the effects of parton showering and calorimeter
segmentation.  A more complete analysis would study the shaping of the 
distributions from the top tagging method using in selecting a sample,
but we expect these effects to be small.

Boosted hadronic tops may provide a new window to shed light on otherwise difficult 
aspects of new physics at the LHC, and will certainly provide a complementary probe 
of physics beyond the Standard Model.  Variables which can analyze
the polarization of boosted hadronic tops, such as those introduced here, will fill an important slot in the analysis toolkit
as we try to unravel the physics behind LHC data.

%%%%%%%%%%%%%%%%%%%%%%%%%%%%%%%%%%%%%%%%%%%%%%%%%%%%%%%%
\acknowledgments{The authors would like to thank Gilad Perez, Matt Schwartz, Jesse Thaler, Chris Tully, and Peter Skands for discussions.  The work of J.S. was supported in part by the DOE grants DE-FG02-96ER40949 and DE-FG02-92ER40704.  The work of L.-T. W. was supported by NSF grant PHY-0756966 and DOE grant DE-FG02-90ER40542.}

\appendix
%%%%%%%%%%%%%%%%%%%%%%%%%%%%%%%%%%%%%%%%%%%%%%%%%%%%%%%%
\section{Angular Distributions in Decays of Polarized Tops}
%%%%%%%%%%%%%%%%%%%%%%%%%%%%%%%%%%%%%%%%%%%%%%%%%%%%%%%%

Here we collect some results on the energy and angular distributions of daughter
partons in polarized top decay.  To arrive at these results we assume the standard model top decay $t\to W b$ with 
subsequent $W $ decay $W\to \bar d u$, and work at leading order in the narrow width approximation for
both $t$ and $W$.  Then, the squared matrix element for a polarized top can be written 
\be
| \mathcal{M} | ^ 2 \propto (t_\pm \cdot d) (b\cdot u)
\ee
where $d,\, b,$ and $u$ denote the momentum four-vectors of the respective partons, and
$t_\pm \equiv p_t \pm m_t S$ contains information about the top polarization through the
spin four-vector $S$.  In the top rest frame, the spin four-vector takes the form $S ^\mu
= (0, \hat s)$, where $\hat s$ is a unit vector defining the axis of polarization.
In the narrow width approximation, and further taking $m_b=0$, the energy of the $b$
quark is fixed at $E_b = (m_t ^ 2-m_W ^ 2)/2 m_t$ in the top rest frame.  The full differential
decay rate then depends nontrivially on two quantities, which we can take to be the
down-quark energy in the top rest frame, $E_d$ and the angle of the down quark with respect to
the top spin axis, $\cos \theta_d$.  The differential decay rate can be written
\be
\frac{1}{\Gamma}\frac{d\Gamma}{d E_d d\cos\theta_d} = 
      \frac{12 m_t ^ 3}{ m_t ^ 6 + 2 m_W^6-3m_W^4 m_t^2}\; E_d (m_t -2 E_d )
          (1+ \mathcal{P}_t \cos \theta_d)
\ee
where $\mathcal{P}_t$ is the top polarization, $-1\leq \mathcal{P}_t \leq 1$ with $\mathcal{P}_t=1,-1$ corresponding to right and left handed tops, respectively.

These results can be generalized to study the angular distribution of any given daughter parton $i$ in the top rest frame by writing~\cite{Jezabek:1988ja}
\be
\frac{1}{\Gamma}\frac{d\Gamma}{ d\cos\theta_i} = {1\over 2} (1+ \mathcal{P}_t\kappa_i \cos \theta_i),
\ee
where $\kappa_i$ is the {\it spin analyzing power} of the parton $i$,
and $\cos\theta_i$ is the angle that parton makes with respect to the top spin
axis in the top rest frame.  The values of $\kappa_i$ for various
choices of $i$ are listed in Tab.~\ref{table:awesomeanalyzingpower}.
The $d $, which corresponds to the lepton in leptonic top decays, is
maximally correlated with the top spin, with $\kappa_d = 1$.  In
addition to the partonic $b$, $W$, $u$, and $d$, we also consider an
object $j$, defined to be the softer of the two light quark jets in
the top rest frame.  As the $d$ tends to be softer than the $u$ in the
top rest frame, this jet is the $d$-jet approximately 60\% of the time
\cite{Jezabek:1994qs,Brandenburg:2002xr}.

%%%%%%%%%%%%%%%%%%%%%%%%%%%%%%%%%%%%%%%%%%%%%%%%%%%%%%
%%% Spin analyzing power
%%%%%%%%%%%%%%%%%%%%%%%%%%%%%%%%%%%%%%%%%%%%%%%%%%%%%%
%\begin{table}
%\begin{center}
%\begin{tabular}{cc}
%\hline \hline
%& \\

%Parton &  $\kappa$  \\
%&  \\
% \hline
%&  \\

%$b $ &  $ -0.4$\\
%$W $ &  $0.4$ \\
%$d $ & 1 \\
%$u $ & $-0.3$ \\
%$j $ & 0.5 \\
%& \\
%\hline \hline
%\end{tabular}

\TABLE[t]{
\parbox{\textwidth}{
%\begin{table}
\begin{center}
\begin{tabular}{|c||ccccc|}
\hline
Parton & $b$ & $W$& $d$ & $u$ & $j$\\
\hline 
 $\kappa$  &$-0.4$& $0.4$ & $1.0$ & $-0.3$ & $0.5$ \\
 \hline
\end{tabular}
\end{center}
\caption{Tree level values for the spin analyzing power $\kappa$ of various 
top daughters in top decay.  The object $j$ is defined in the text.
\label{table:awesomeanalyzingpower}
}
}
}
%\end{table}
%%%%%%%%%%%%%%%%%%%%%%%%%%%%%%%%%%%%%%%%%%%%%%%%%%%%%%%%

\bibliography{chiraltop}

\providecommand{\href}[2]{#2}\begingroup\raggedright\begin{thebibliography}{10}

\bibitem{Dimopoulos:1981zb}
S.~Dimopoulos and H.~Georgi, {\it {Softly Broken Supersymmetry and SU(5)}},
  {\em Nucl. Phys.} {\bf B193} (1981) 150.

\bibitem{ArkaniHamed:2001nc}
N.~Arkani-Hamed, A.~G. Cohen, and H.~Georgi, {\it {Electroweak symmetry
  breaking from dimensional deconstruction}},  {\em Phys. Lett.} {\bf B513}
  (2001) 232--240, [\href{http://xxx.lanl.gov/abs/hep-ph/0105239}{{\tt
  hep-ph/0105239}}].

\bibitem{Agashe:2003zs}
K.~Agashe, A.~Delgado, M.~J. May, and R.~Sundrum, {\it {RS1, custodial isospin
  and precision tests}},  {\em JHEP} {\bf 08} (2003) 050,
  [\href{http://xxx.lanl.gov/abs/hep-ph/0308036}{{\tt hep-ph/0308036}}].

\bibitem{Kane:1991bg}
G.~L. Kane, G.~A. Ladinsky, and C.~P. Yuan, {\it {Using the top quark for
  testing standard model polarization and CP predictions}},  {\em Phys. Rev.}
  {\bf D45} (1992) 124--141.

\bibitem{Agashe:2006hk}
K.~Agashe, A.~Belyaev, T.~Krupovnickas, G.~Perez, and J.~Virzi, {\it {LHC
  signals from warped extra dimensions}},  {\em Phys. Rev.} {\bf D77} (2008)
  015003, [\href{http://xxx.lanl.gov/abs/hep-ph/0612015}{{\tt
  hep-ph/0612015}}].

\bibitem{Lillie:2007yh}
B.~Lillie, L.~Randall, and L.-T. Wang, {\it {The Bulk RS KK-gluon at the LHC}},
   {\em JHEP} {\bf 09} (2007) 074,
  [\href{http://xxx.lanl.gov/abs/hep-ph/0701166}{{\tt hep-ph/0701166}}].

\bibitem{Barger:2006hm}
V.~Barger, T.~Han, and D.~G.~E. Walker, {\it {Top Quark Pairs at High Invariant
  Mass: A Model- Independent Discriminator of New Physics at the LHC}},  {\em
  Phys. Rev. Lett.} {\bf 100} (2008) 031801,
  [\href{http://xxx.lanl.gov/abs/hep-ph/0612016}{{\tt hep-ph/0612016}}].

\bibitem{Fitzpatrick:2007qr}
A.~L. Fitzpatrick, J.~Kaplan, L.~Randall, and L.-T. Wang, {\it {Searching for
  the Kaluza-Klein Graviton in Bulk RS Models}},  {\em JHEP} {\bf 09} (2007)
  013, [\href{http://xxx.lanl.gov/abs/hep-ph/0701150}{{\tt hep-ph/0701150}}].

\bibitem{Skiba:2007fw}
W.~Skiba and D.~Tucker-Smith, {\it {Using jet mass to discover vector quarks at
  the LHC}},  {\em Phys. Rev.} {\bf D75} (2007) 115010,
  [\href{http://xxx.lanl.gov/abs/hep-ph/0701247}{{\tt hep-ph/0701247}}].

\bibitem{Baur:2007ck}
U.~Baur and L.~H. Orr, {\it {High $p_{T}$ Top Quarks at the Large Hadron
  Collider}},  {\em Phys. Rev.} {\bf D76} (2007) 094012,
  [\href{http://xxx.lanl.gov/abs/0707.2066}{{\tt 0707.2066}}].

\bibitem{Frederix:2007gi}
R.~Frederix and F.~Maltoni, {\it {Top pair invariant mass distribution: a
  window on new physics}},  {\em JHEP} {\bf 01} (2009) 047,
  [\href{http://xxx.lanl.gov/abs/0712.2355}{{\tt 0712.2355}}].

\bibitem{Baur:2008uv}
U.~Baur and L.~H. Orr, {\it {Searching for $t \bar{t}$ Resonances at the Large
  Hadron Collider}},  {\em Phys. Rev.} {\bf D77} (2008) 114001,
  [\href{http://xxx.lanl.gov/abs/0803.1160}{{\tt 0803.1160}}].

\bibitem{Thaler:2008ju}
J.~Thaler and L.-T. Wang, {\it {Strategies to Identify Boosted Tops}},  {\em
  JHEP} {\bf 07} (2008) 092, [\href{http://xxx.lanl.gov/abs/0806.0023}{{\tt
  0806.0023}}].

\bibitem{Kaplan:2008ie}
D.~E. Kaplan, K.~Rehermann, M.~D. Schwartz, and B.~Tweedie, {\it {Top Tagging:
  A Method for Identifying Boosted Hadronically Decaying Top Quarks}},  {\em
  Phys. Rev. Lett.} {\bf 101} (2008) 142001,
  [\href{http://xxx.lanl.gov/abs/0806.0848}{{\tt 0806.0848}}].

\bibitem{Almeida:2008yp}
L.~G. Almeida {\em et~al.}, {\it {Substructure of high-$p_T$ Jets at the LHC}},
   {\em Phys. Rev.} {\bf D79} (2009) 074017,
  [\href{http://xxx.lanl.gov/abs/0807.0234}{{\tt 0807.0234}}].

\bibitem{Almeida:2008tp}
L.~G. Almeida, S.~J. Lee, G.~Perez, I.~Sung, and J.~Virzi, {\it {Top Jets at
  the LHC}},  {\em Phys. Rev.} {\bf D79} (2009) 074012,
  [\href{http://xxx.lanl.gov/abs/0810.0934}{{\tt 0810.0934}}].

\bibitem{Bai:2008sk}
Y.~Bai and Z.~Han, {\it {Top-antitop and Top-top Resonances in the Dilepton
  Channel at the CERN LHC}},  {\em JHEP} {\bf 04} (2009) 056,
  [\href{http://xxx.lanl.gov/abs/0809.4487}{{\tt 0809.4487}}].

\bibitem{cmsca}
T.~C. Collaboration, {\it A cambridge-aachen (c-a) based jet algorithm for
  boosted top-jet tagging},  Tech. Rep. CMS PAS JME-09-001, Jul, 2007.

\bibitem{Brooijmans}
G.~Brooijmans, {\it High $p_t$ hadronic top quark identification},  Tech. Rep.
  ATL-COM-PHYS-2008-001, ATLAS, Feb, 2008.

\bibitem{Ellis:2009su}
S.~D. Ellis, C.~K. Vermilion, and J.~R. Walsh, {\it {Techniques for improved
  heavy particle searches with jet substructure}},
  \href{http://xxx.lanl.gov/abs/0903.5081}{{\tt 0903.5081}}.

\bibitem{Shelton:2008nq}
J.~Shelton, {\it {Polarized tops from new physics: signals and observables}},
  {\em Phys. Rev.} {\bf D79} (2009) 014032,
  [\href{http://xxx.lanl.gov/abs/0811.0569}{{\tt 0811.0569}}].

\bibitem{Perelstein:2008zt}
M.~Perelstein and A.~Weiler, {\it {Polarized Tops from Stop Decays at the
  LHC}},  {\em JHEP} {\bf 03} (2009) 141,
  [\href{http://xxx.lanl.gov/abs/0811.1024}{{\tt 0811.1024}}].

\bibitem{March:851053}
L.~March, E.~Ros, and B.~Salvachœa, {\it Search for kaluza-klein excitations of
  the gluon in models with extra dimensions},  Tech. Rep.
  ATL-PHYS-PUB-2006-002. ATL-COM-PHYS-2005-032, CERN, Geneva, Jul, 2005.

\bibitem{Gonz‡lezdelaHoz:814346}
S.~Gonz‡lez de~la Hoz, L.~March, and E.~Ros, {\it Search for hadronic decays of
  $z_{H}$ and $w_{H}$ in the little higgs model},  Tech. Rep.
  ATL-PHYS-PUB-2006-003. ATL-COM-PHYS-2005-001, CERN, Geneva, 2005.

\bibitem{Lehmacher:2008hs}
M.~Lehmacher, {\it {b-Tagging Algorithms and their Performance at ATLAS}},
  \href{http://xxx.lanl.gov/abs/0809.4896}{{\tt 0809.4896}}.

\bibitem{Cacciari:Fastjet}
M.~Cacciari, G.~Salam, and G.~Soyez, ``{FastJet}.'' http://fastjet.fr/.

\bibitem{Maltoni:2002qb}
F.~Maltoni and T.~Stelzer, {\it {MadEvent: Automatic event generation with
  MadGraph}},  {\em JHEP} {\bf 02} (2003) 027,
  [\href{http://xxx.lanl.gov/abs/hep-ph/0208156}{{\tt hep-ph/0208156}}].

\bibitem{Sjostrand:2006za}
T.~Sjostrand, S.~Mrenna, and P.~Skands, {\it {PYTHIA 6.4 physics and manual}},
  {\em JHEP} {\bf 05} (2006) 026,
  [\href{http://xxx.lanl.gov/abs/hep-ph/0603175}{{\tt hep-ph/0603175}}].

\bibitem{Bahr:2008pv}
M.~Bahr {\em et~al.}, {\it {Herwig++ Physics and Manual}},  {\em Eur. Phys. J.}
  {\bf C58} (2008) 639--707, [\href{http://xxx.lanl.gov/abs/0803.0883}{{\tt
  0803.0883}}].

\bibitem{Cacciari:2008gp}
M.~Cacciari, G.~P. Salam, and G.~Soyez, {\it {The anti-$k_t$ jet clustering
  algorithm}},  {\em JHEP} {\bf 04} (2008) 063,
  [\href{http://xxx.lanl.gov/abs/0802.1189}{{\tt 0802.1189}}].

\bibitem{Jezabek:1988ja}
M.~Jezabek and J.~H. Kuhn, {\it {Lepton Spectra from Heavy Quark Decay}},  {\em
  Nucl. Phys.} {\bf B320} (1989) 20.

\bibitem{Jezabek:1994qs}
M.~Jezabek, {\it {Top quark physics}},  {\em Nucl. Phys. Proc. Suppl.} {\bf
  37B} (1994) 197, [\href{http://xxx.lanl.gov/abs/hep-ph/9406411}{{\tt
  hep-ph/9406411}}].

\bibitem{Brandenburg:2002xr}
A.~Brandenburg, Z.~G. Si, and P.~Uwer, {\it {QCD-corrected spin analysing power
  of jets in decays of polarized top quarks}},  {\em Phys. Lett.} {\bf B539}
  (2002) 235--241, [\href{http://xxx.lanl.gov/abs/hep-ph/0205023}{{\tt
  hep-ph/0205023}}].

\end{thebibliography}\endgroup
\bibliographystyle{jhep}
\end{document}